%% file: paper-arXiv.tex
\documentclass[sigconf,nonacm]{acmart}
\AtBeginDocument{%
  }

\setcopyright{none}
\renewcommand\footnotetextcopyrightpermission[1]{}
\usepackage[english]{babel}
\usepackage{enumitem}
\usepackage{amsmath}
\usepackage{amsthm}
\usepackage{graphicx}
\usepackage{ragged2e}
\usepackage{multirow}
\usepackage{bm} 
\usepackage{url}
\usepackage{hyperref}
\usepackage{algorithm,algorithmicx}
\usepackage[noend]{algpseudocode}
\usepackage{color, soul}
\usepackage{indentfirst}
\usepackage{caption,subcaption}
\usepackage{listings} 
\usepackage[flushleft]{threeparttable}
\usepackage{tabularx,colortbl}
\usepackage{booktabs}
\usepackage{tikz}
\usepackage{pifont}

\usepackage{amssymb}
\usepackage{xspace}

\definecolor{OliveGreen}{cmyk}{0.64,0,0.95,0.40}
\definecolor{ao}{rgb}{0.0, 0.5, 0.0}
\definecolor{asparagus}{rgb}{0.53, 0.66, 0.42}
\definecolor{applegreen}{rgb}{0.55, 0.71, 0.0}
\definecolor{aogreen}{rgb}{0.0, 0.5, 0.0}
\definecolor{columbiablue}{rgb}{0.61, 0.87, 1.0}
\definecolor{cornellred}{rgb}{0.7, 0.11, 0.11}
\definecolor{cornflowerblue}{rgb}{0.39, 0.58, 0.93}
\definecolor{denim}{rgb}{0.08, 0.38, 0.74}

\colorlet{BtfulGreen}{black!30!green!70!}
\colorlet{BtfulOrange}{white!10!orange!90!}
\colorlet{BtfulGray}{white!50!gray!50!}

\newcommand{\cut}[1]{}

\newcommand{\sysname}{OmniPlan\xspace}

\setlist{nolistsep}
\begin{document}

\title{\sysname: An Adaptive Framework for Timely and Near-Optimal Network Planning Optimization}

\author{Longlong Zhu}
\orcid{0000-0001-6128-3065}
\affiliation{
 \institution{Zhejiang University}
 \city{Hangzhou}
 \state{Zhejiang}
 \country{China}
}
\email{kikionepin@gmail.com}

\author{Jiashuo Yu}
\orcid{0009-0001-9056-8534}
\affiliation{
 \institution{Zhejiang University}
 \city{Hangzhou}
 \state{Zhejiang}
 \country{China}
}
\email{joshuayu.2048@gmail.com}

\author{Zedi Chen}
\orcid{0009-0002-5346-9577}
\affiliation{
 \institution{Zhejiang University}
 \city{Hangzhou}
 \state{Zhejiang}
 \country{China}
}
\email{chenzedi@zju.edu.cn}

\author{Yuhan Wu}
\orcid{0009-0003-8364-1611}
\affiliation{
 \department{College of Computer and Data Science}
 \institution{Fuzhou University}
 \city{Fuzhou}
 \state{Fujian}
 \country{China}
}
\email{yharim.io@gmail.com}

\author{Zhifan Jiang}
\orcid{0009-0008-0212-9330}
\affiliation{
 \institution{Zhejiang University}
 \city{Hangzhou}
 \state{Zhejiang}
 \country{China}
}
\email{jiangzf1231@zju.edu.cn}

\author{Yuchen Xian}
\orcid{0009-0005-0254-8365}
\affiliation{
 \institution{Zhejiang University}
 \city{Hangzhou}
 \state{Zhejiang}
 \country{China}
}
\email{yuchen_xian@zju.edu.cn}

\author{Yimeng Liu}
\orcid{0009-0009-7976-6079}
\affiliation{
 \institution{Zhejiang University}
 \city{Hangzhou}
 \state{Zhejiang}
 \country{China}
}
\email{liuyimeng@zju.edu.cn}

\author{Jiajie Su}
\orcid{0000-0002-6899-4174}
\affiliation{
 \institution{Zhejiang University}
 \city{Hangzhou}
 \state{Zhejiang}
 \country{China}
}
\email{sujiajie@zju.edu.cn}

\author{Shaopeng Zhou}
\orcid{0009-0007-7737-6283}
\affiliation{
 \institution{Zhejiang University}
 \city{Hangzhou}
 \state{Zhejiang}
 \country{China}
}
\email{abnerzhou@zju.edu.cn}

\author{Xingyuan Li}
\orcid{0000-0001-9081-817X}
\affiliation{
 \institution{Zhejiang University}
 \city{Hangzhou}
 \state{Zhejiang}
 \country{China}
}
\email{xingyuan_lxy@163.com}

\author{Hongyan Liu}
\orcid{0000-0001-7753-0294}
\affiliation{
 \institution{Fuzhou University}
 \city{Fuzhou}
 \state{Fujian}
 \country{China}
}
\email{hyliu20@zju.edu.cn}

\author{Xuan Liu}
\orcid{0000-0002-7966-4488}
\affiliation{
 \institution{Yangzhou University}
 \city{Yangzhou}
 \state{Jiangsu}
 \country{China}
}
\email{yusuf@yzu.edu.cn}

\author{Dong Zhang}
\orcid{0000-0002-6379-0244}
\affiliation{
 \department{College of Computer and Data Science}
 \institution{Fuzhou University}
 \city{Fuzhou}
 \state{Fujian}
 \country{China}
}
\email{zhangdong@fzu.edu.cn}

\author{Chunming Wu}
\orcid{0000-0001-7958-9687}
\affiliation{
 \institution{Zhejiang University}
 \city{Hangzhou}
 \state{Zhejiang}
 \country{China}
}
\email{wuchunming@zju.edu.cn}

\author{Xiang Chen}
\orcid{0000-0002-0249-9664}
\authornote{Xiang Chen is the corresponding author.\\[3pt]
\textit{Accepted by ACM KDD 2026.}}
\affiliation{
 \department[1]{The State Key Laboratory of Blockchain and Data Security}
 \department[0]{College of Computer Science and Technology}
 \institution{Zhejiang University}
 \city{Hangzhou}
 \state{Zhejiang}
 \country{China}
}
\email{wasdnsxchen@gmail.com}

\renewcommand{\shortauthors}{Longlong Zhu et al.}

\begin{abstract}
\noindent Network planning optimization is a fundamental problem across diverse domains, including transportation systems, communication networks, and power grids. It requires simultaneous optimization of multiple competing objectives under complex constraints. Existing network planning optimization frameworks rely on mixed integer programming (MIP) solvers, heuristics, and deep reinforcement learning (DRL) models to compute planning decisions. However, they lack effective adaptability to diverse and dynamic user intents, thus leading to the trade-off between execution time and optimality. In this paper, we propose OmniPlan, an adaptive framework that achieves both timeliness and near-optimality in network planning optimization. To achieve the adaptability lacking in existing solutions, OmniPlan employs a large language model (LLM)-based interpreter to convert heterogeneous natural-language intents into a unified and quantifiable user-preference vector. Then it employs a mixture-of-experts architecture that integrates MIP solvers, heuristics, and DRL models as specialized experts, where OmniPlan adapts to diverse intents by dynamically selecting timely and near-optimal experts. Finally, it incorporates a DRL-based expert configuration module that fine-tunes optimization objective weights to align planning decisions with user-specific preferences. We evaluate OmniPlan with a representative real-world workload, i.e., distributed machine learning (ML), where we leverage OmniPlan to offload a wide spectrum of ML inference tasks, e.g., decision trees, SVM, naive Bayes, XGBoost, and random forests, onto a network of hardware devices. Our experiments on a real-world testbed indicate that OmniPlan achieves near-optimal and low-execution-time offloading for real-world ML inference tasks, reducing latency by up to 97.8\% and network device resource consumption by up to 11.5\%.
\end{abstract}


\maketitle

\input{paper/introduction}
\input{paper/background}
\input{paper/overview}
\input{paper/design1}
\input{paper/design2}

\input{paper/design3}

\input{paper/evaluation}
\input{paper/discussion}
\input{paper/limitations}
\input{paper/conclusion}

\section*{Acknowledgments}
\noindent This work is supported by the National Key R\&D Program of China (2024YFB2906503), the National Natural Science Foundation of China (Grant No. 62602005), the Science and Technology Program of Fujian Province (Grant No. 2025H6007), and the Zhejiang Xinmiao Talents Program (Grant No. 2026R401172).

\bibliographystyle{ACM-Reference-Format}
\balance
\bibliography{paper}

\clearpage

\input{paper/appendix}

\end{document}

%% file: paper/introduction.tex
\section{Introduction}
\noindent Network planning optimization is fundamental to infrastructure systems, e.g., transportation networks, communication systems, and power grids \cite{Gu2025LLMsWorkflow115,Xu2024AIGCNetworks99,du2024mixture,ruan2023tptulargelanguagemodelbased159}. It requires simultaneous optimization of multiple competing objectives from the user intent, e.g., minimizing delay, maximizing throughput, and minimizing resource usage, under complex constraints. 
For example, in communication networks, operators need to deploy services across distributed network devices to minimize end-to-end latency while reducing resource usage, with respect to device resource capacity and link bandwidth constraints \cite{liu2024speed, chen2024eagle}. 
To achieve near-optimal planning, existing network planning optimization frameworks typically leverage mixed integer programming (MIP) solvers, heuristics, and deep reinforcement learning (DRL) models to compute planning decisions.

However, prior studies suffer from the trade-off between execution time and optimality, leading to either long execution time or significant optimality loss. 
(1) MIP-based approaches formulate network planning optimization as an MIP problem \cite{jose2015compiling,gupta2018sonata,sultana2021flightplan,gao2020lyra,hogan2022modular} and leverage off-the-shelf MIP solvers to compute optimal planning decisions. 
However, they suffer from a long execution time to search for new decisions when the input task changes due to the need to enumerate points within the large solution space. 
(2) Heuristic-based approaches \cite{liu2024speed,chen2024hermes,chen2024eagle} introduce dedicated heuristics for computing near-optimal planning decisions in polynomial time.
However, they suffer from significant optimality loss when the input task changes since heuristics are designed for specific optimization tasks.
(3) DRL model-based approaches achieve near-optimal network planning for different tasks \cite{zhang2022efficient,pei2018efficiently,zhang2023dapper}. Here, they meet the need of dynamically computing planning decisions for new incoming tasks via trained DRL agents with respect to the specific user intent.
However, they suffer from significant optimality loss when user intent changes since they are typically built for specific user intents. Otherwise, they require time-consuming retraining or redesign to support new user intents. 
At the core, existing solutions fall short because they fail to adapt to changing user intents, facing the trade-off between execution time and optimality.
In response, we observe that the powerful semantic understanding capability of large language models (LLMs) and the dynamic adaptability in handling heterogeneous tasks of the mixture-of-experts (MoE) architecture can offer an opportunity to address this trade-off. 
More precisely, LLMs exhibit strong semantic understanding capabilities \cite{wu2024netllm,zhang2022opt} that enable them to comprehend and reason about diverse user intents expressed in natural languages. 
Furthermore, MoE that combines multiple experts and selectively activates the most suitable experts for each input \cite{masoudnia2014mixture} has been widely applied across numerous applications, such as network planning optimization \cite{du2024mixture}, natural language processing \cite{zhuang2024litemoe,huang2024mc}, and multi-modal learning \cite{mustafa2022multimodal,xue2023raphael,wu2024robust}.
MoE shows exceptional adaptability to different optimization objectives and can activate suitable experts for specific input user intent.

In this paper, we propose \sysname, an adaptive framework for timely and near-optimal network planning optimization. 
To address the lack of adaptability to diverse and dynamic intents, \sysname leverages LLMs to understand user intents, based on which it activates timely and near-optimal experts within the MoE architecture.
More precisely, \sysname first interprets heterogeneous natural language intents of network planning into a unified and quantifiable user preference vector via LLMs. 
Second, \sysname adapts to different intents by activating timely and near-optimal experts. 
It selectively activates fast DRL/heuristic-based experts that compute near-optimal planning decisions and optimal solver-based experts that are pre-configured for optimizing specific intents. 
Third, \sysname tailors planning decisions to the specific user intent by configuring weights of optimization objectives in the activated expert. 
Here, values within the preference vector cannot be directly applied as weights in solvers due to dimensional heterogeneity, so \sysname dynamically adjusts weights of optimization objectives via intent-aware DRL-based expert configurations.

\noindent \textbf{Contributions}. We make the following contributions.
\begin{itemize}[leftmargin=*]
    \item We propose \sysname, an adaptive framework for timely and near-optimal network planning optimization. It converts natural language intents into a quantifiable preference vector via LLMs. 
    \item We propose a dual-channel activation approach that activates fast DRL or heuristic experts for near-optimal planning and optimal MIP experts for a specific set of optimization objectives.
    \item We develop a DRL-based expert configuration approach that dynamically adjusts weights within the activated experts, tailoring decisions to specific user intents.
    \item We implement the \sysname prototype and evaluate it on a representative real-world workload, i.e., distributed ML, where we leverage \sysname to offload a wide spectrum of ML inference tasks \cite{zheng_dinc_2023} onto a network of hardware devices. The experimental results show that \sysname achieves near-optimal and low-execution-time offloading for real-world ML inference tasks, reducing network device resource consumption by up to 11.5\% and latency by up to 97.8\%.
\end{itemize}

%% file: paper/background.tex
\section{Background and Motivation}

\subsection{Existing Works and Limitations}
\noindent Existing network planning works can be typed into three classes, including MIP solvers, heuristics, and DRL model-based solutions.

\noindent \textbf{MIP solvers-based solutions} \cite{liu2020sra,chen2020speed,chen2021mtp,sultana2021flightplan,xu2023clickinc,gao2020lyra,hogan2022modular} formulate the network planning task as an MIP problem, preserving optimality at the cost of long execution time. They leverage MIP solvers like Gurobi \cite{gurobi} and CPLEX \cite{cplex} to compute optimal planning decisions. 
For example, SPEED \cite{liu2024speed} formulates the resource allocation task in a communication system as an MIP problem, aiming at minimizing resource consumption and end-to-end latency. 
However, they consume high computational overhead for enumerating points within the large solution space, which leads to extensive execution time.

\noindent \textbf{Heuristics.} Heuristic-based solutions \cite{chen2024hermes,chen2024eagle,liu2024speed} compute planning decisions in time, but suffer from optimality loss. 
They develop dedicated heuristic algorithms, such as greedy-based, rounding-based, and Bender's decomposition-based algorithms, for specific planning tasks. Thus, they can obtain near-optimal planning decisions in polynomial time. For example, Hermes \cite{chen2024hermes} develops a greedy-based task partitioning heuristic for minimizing coordination overhead across network nodes. 
However, heuristics are tailored for specific tasks, which results in optimality loss when user intents change.

\noindent \textbf{DRL models.} DRL models cannot navigate the trade-off between execution time and decision quality when input tasks change. 
DRL models \cite{zhang2022efficient,pei2018efficiently,zhang2023dapper} are typically used for service orchestration. For example, Dapper \cite{zhang2023dapper} combines DRL and graph convolutional networks to compute near-optimal orchestration decisions in time. 
Here, we can leverage the adaptive learning capabilities of DRL models to meet the need of dynamically computing planning decisions for new incoming tasks with respect to a specific intent. 
However, they need to be retrained or even redesigned for different types of user intents; otherwise, they suffer from high optimality loss. 
In this case, the retraining or redesigning requires a high engineering cost since their performance depends on model designs, while constructing complex DRL models is labor-intensive \cite{wu2024netllm,du2024mixture}. For example, deep neural networks (DNNs) contain thousands to billions of neurons, which require day-level tuning time \cite{wu2024netllm}.

\subsection{Opportunities of LLMs and MoE}
\noindent MoE is a neural network architecture that combines multiple experts with a gateway network to selectively activate the most appropriate experts for each input \cite{masoudnia2014mixture}. It offers significant advantages in terms of parameter and model scaling efficiency by activating only a subset of parameters for each inference \cite{yuksel2012twenty,vats2024theEvolution}. Thus, it has been widely applied across diverse domains, such as network planning optimization \cite{du2024mixture}, natural language processing \cite{zhuang2024litemoe,huang2024mc}, computer vision \cite{gross2017hard,li2024sm3det,xu2025limoe}, and multi-modal learning \cite{mustafa2022multimodal,xue2023raphael,wu2024robust}.

Inspired by MoE's successes, we explore its application in network planning optimization. It brings the opportunity to address the challenge of intent-adaptive near-optimal planning \cite{du2023user,du2024mixture}. 
In detail, it can integrate multiple specialized planning experts, each optimized for a specific task. It employs a gating network to selectively activate a few experts tailored for specific inputs. Thus, it allows for dynamic expert activation based on user intents.

Moreover, LLMs offer a promising approach that achieves intent-aware network planning thanks to their semantic understanding and timely decision-making capabilities \cite{wu2024netllm,zhang2022opt}. 
Compared to traditional planning approaches, LLMs can comprehend complex planning requirements expressed in natural languages, e.g., interpreting user intents and translating them into specific optimization objectives. This semantic understanding enables LLMs to dynamically adjust optimization priorities based on specific contexts and network dynamics, allowing them to achieve intent-aware planning.

By combining LLMs with specialized experts via an MoE architecture, we can achieve both near-optimal and intent-aware network planning. 
In particular, such a combination allows us to avoid designing a new specialized DRL model or heuristic for each planning task or intent, which significantly reduces engineering costs while maintaining high-quality planning decisions. 
Furthermore, we integrate diverse types of planning experts, including but not limited to DRL-based models, and use the LLM as an orchestrator that interprets user intents and activates the most suitable expert(s) for the given planning context.

\subsection{LLMs and MoE for Network Planning}
\noindent Recent studies apply LLMs and MoE to network planning tasks.
On the LLM side, they explore network service placement \cite{li2025llmSketch}, traffic management \cite{chattopadhyay2022mixture,jiang2024interpretable}, and job scheduling \cite{wu2024netllm}. 
On the MoE side, the MoE architecture is applied to LLM-assisted network planning applications. It enables reducing the need for task-specific model training \cite{du2024mixture} and deploying experts on resource-constrained network infrastructures, such as access points, edge servers, and mobile devices \cite{xu2025decentralization}. 
These works address different aspects and can be integrated with \sysname in a complementary manner.

%% file: paper/overview.tex
\begin{figure*}[t]
    \centering
    \includegraphics[width=0.95\linewidth]{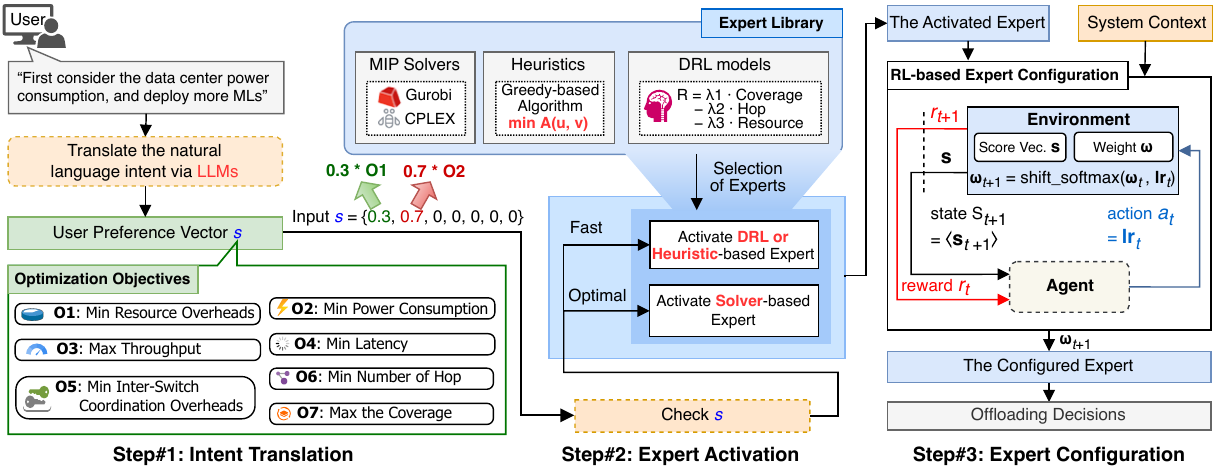}
    \caption{The architecture of OmniPlan.}
    \label{fig:arc}
\end{figure*}

\section{Overview}\label{methodology}
\noindent We propose \sysname to break the trade-off between execution time and optimality under heterogeneous intents.

\noindent \textbf{Key ideas.} \sysname aims to achieve two goals:
(1) G1: \textit{optimal or near-optimal} and \textit{fast} planning that achieves low execution time while maintaining solution quality, and (2) G2: \textit{intent-aware} planning that computes near-optimal planning decisions under different user intents without redesigning and retraining.
To achieve these goals, the key idea of \sysname is to leverage LLMs to understand user intents and activate suitable experts within the MoE architecture to achieve intent-adaptive planning. Here, LLMs understand a wide variety of natural language input user intents (\textbf{achieving G2}) and MoE dynamically activates suitable experts for given intents without redesigning and retraining (\textbf{achieving G1}).

\noindent \textbf{\sysname workflow.} \sysname executes three major steps.

\textit{Step\#1: Intent translation.} 
\sysname interprets heterogeneous input user intents described in natural language into a unified and quantifiable preference vector $s$ via LLMs, where $s$ is a normalized preference vector consisting of some optimization objectives. 
By $s$, \sysname captures optimization priorities within input intents.

\textit{Step\#2: Expert activation.} 
\sysname activates suitable experts based on the interpreted vector $s$ and expert scores. For intents where the weight of a specific objective within $s$ exceeds the threshold $\tau$ (e.g., $\omega_{\text{lat}} \ge 0.9$), \sysname \textbf{only} activates a DRL model or heuristic-based expert that computes near-optimal planning decisions in time. 
Otherwise, in addition to the selected expert for timely planning decisions, \sysname activates a solver-based expert that computes optimal planning decisions. 
Here, the speedup of \sysname comes from avoiding expensive solvers when a near-optimal heuristic suffices. 
We develop DRL model-based and heuristic-based experts for different categories of planning intents, including node-level, link-level, and network-wide intents.
We also develop MIP solver-based experts with typical optimization objective configurations. For example, we can build an expert with a configuration that minimizes the weighted sum of resource overhead and latency, with optimized weights assigned to each objective.

\textit{Step\#3: Expert configuration.}  
For the activated expert in Step\#2, \sysname configures it to fit the user intent. More precisely, values within the preference vector cannot be directly applied as weights in solvers, because solvers cannot normalize performance metrics, and direct adoption distorts optimization outcomes as magnitude differences dominate user intent.
Thus, \sysname employs an intent-aware DRL-based expert configuration approach to tune weights within the optimization objectives in the activated solver-based expert.
Here, \sysname develops an intent-aware scoring approach that maps the non-uniformly distributed intent space to a uniform scoring space, where \sysname quantifies the alignment between expert configurations and user intents. Then, the DRL agent integrates this intent-aware score into its state space and reward function to fine-tune optimization objective weights for achieving better performance for the input intent.

%% file: paper/design1.tex
\section{Intent Translation}\label{intentInterpreting} 
\noindent \textbf{Challenges.} The network planning task needs to handle diverse optimization objectives arising from heterogeneous task requirements, such as minimizing latency for time-sensitive tasks and maximizing throughput for data-intensive tasks. 
Here, user intents may contain multiple objectives. For example, the intent of ``offloading more tasks in a green data center'' involves minimizing both hardware resource overhead and power consumption.
Consequently, the user intent exhibits significant diversity since the objective combination is varied. 
Meanwhile, it is challenging to quantify user evaluations of varying degrees semantically and transform them into evaluation scores that are uniformly distributed between 0 and 1. For example, the model must differentiate the subtle nuances between semantically similar terms like ``high'' and ``great'', while capturing the vast disparity between opposing terms such as ``little'' and ``greatest''.

\noindent \textbf{Key ideas.} To handle diverse user planning intents, we build three fundamental intent categories.
Then, we use an LLM-powered intent translation layer to convert user natural language into standardized textual representations.
Finally, we propose an intent-aware scoring approach that maps intents, which are non-uniformly distributed in the textual space, to a quantified scoring space.

\noindent \textbf{Fundamental intents identification.} 
We summarized that most of the network planning intents can be decomposed into three fundamental intent classes.
We identify these types of intents based on their optimization scope. 
(1) \textit{Node-level intents} involve minimizing power consumption, resource consumption, and others. 
We formulate this class of intents as $\min \omega_{\text{pwr}}$, and $\min \omega_{\text{res}}$.
(2) \textit{Link-level intents} include maximizing end-to-end performance and reducing inter-device coordination overhead. 
This class is related to optimization objectives that minimize the inter-device overhead and maximize the end-to-end processing performance. 
We formulate this class of intents as $\max \omega_{\text{thr}}$, $\omega_{\text{lat}}$, and $\min \omega_{\text{coor}}$.
(3) \textit{Network-wide intents} refer to optimization objectives that involve minimizing the hop counts, maximizing the coverage rate, and others. 
We formulate these intents as $\min \omega_{\text{hop}}$, and $\max \omega_{\text{cover}}$

\noindent \textbf{Intents translation via LLMs.} We formulate intent translation as a constrained API call to LLMs, where natural-language input $u$ is mapped to standardized textual representations $\mathcal{T}$.
More precisely, we execute a domain-engineered prompt function $\mathcal{P}: \mathcal{U} \to \mathcal{T}$, i.e.,
\begin{equation}
\mathcal{P}(u) = \tau_{\text{net}} + u + \tau_{\text{format}}
\end{equation}
where $\tau_{\text{net}}$ = ``As a network planning optimization expert, retain the description from the user regarding the following specific optimization objectives: power, device resources, throughput, latency, inter-device coordination overhead, hop count, coverage. If no specific requirement is provided, mark the corresponding optimization objective as $default=0$ ''. User intent: ``$u$''. $\tau_{\text{format}}$ = ``Output formats: $[\mathcal{T}_1, \dots, \mathcal{T}_7]$''.

\noindent \textbf{Intent-aware scoring.} It converts the standardized textual representations into scores, comprising four main steps:

\textbf{Step\#1: Intent curve construction.} 
We first define $K$ intent anchor points ${S_0, S_1, \ldots, S_{K-1}}$ that semantically and uniformly describe a gradual progression from ``no demand'' to ``greatest demand''. 
For example, $S_0$ = ``The user has no demand'', $S_1$ = ``The user has a little demand'', and up to $S_4$ = ``The user has the greatest demand''. 
By the pre-defined sentence-BERT model, we encode these anchors into vectors ${p_0, p_1, \ldots, p_{K-1}}$. 
Then, to capture the non-linear distribution of intents in high-dimensional space, we use cubic spline interpolation to construct a smooth curve $\mathcal{L}(\theta)$ between anchor points, where $\theta \in [0, K-1]$ is the curve parameter that satisfies $\mathcal{L}(i) = p_i$.

\textbf{Step\#2: Intent curve mapping.} To map the non-uniform intent curve to a uniform scoring space, we compute the arc-length between adjacent anchor points $L_{i,i+1} = \int_{i}^{i+1} \| \frac{d\mathcal{L}(\theta)}{d\theta} \| d\theta$, and define the arc-length weight for each segment as $\phi_{i,i+1} = \frac{1}{L_{i,i+1}}$. 
Here, regions with large semantic gaps (i.e., long arc-length) are normalized with smaller weights to ensure a uniform scoring progression. By this mapping, we obtain a uniform intent-aware scoring space.

\textbf{Step\#3: Per-objective demand scoring.} 
For each objective $O_i$, given its standardized textual representations $\mathcal{T}_i$, we first compute its embedding vector $p_{u_i} = \text{BERT}(\mathcal{T}_i)$. Then, we find the closest projection point on our intent curve $\mathcal{L}(\theta)$, i.e., $\theta_i^* = \arg\min_{\theta \in [0,K-1]} \|\mathcal{L}(\theta) - p_{u_i}\|$. Finally, its score is calculated through weighted arc-length integration, i.e.,
\begin{equation}
\text{score}(\mathcal{T}_i) = \frac{1}{K-1} \int_0^{\theta_i^*} \phi(\theta) \cdot \left| \frac{d\mathcal{L}(\theta)}{d\theta} \right| d\theta
\end{equation}
which yields a per-objective score vector. These scores are then normalized to ensure $\sum_{i=1}^{M} \omega_i = 1$, producing the final preference weight vector $s \in \mathbf{R}^M$, i.e.,
\begin{equation}
\mathbf{s} = [\omega_{\text{res}}, \omega_{\text{pwr}}, \omega_{\text{thr}}, \omega_{\text{lat}}, \omega_{\text{coor}}, \omega_{\text{hop}}, \omega_{\text{cover}}]
\end{equation}
The vector $\mathbf{s}$ can be extended to contain more than 7 objectives.
By translating input intents into the preference vector $s$, \sysname handles most of the heterogeneous or unknown input intents.

%% file: paper/design2.tex
\section{Expert Activation}
\label{sec:expert_activation}
\noindent \textbf{Challenges.} Given a user intent score vector $\mathbf{s}$, we need to select appropriate experts under user intents. Here, network planning must adapt to dynamic user intents. However, heuristics and DRL-based solutions only handle predefined intent patterns, and MIP solver-based solutions are confined to fixed objective sets.

\noindent \textbf{Key ideas.} We build experts for the above intents. 
Then, \sysname adapts to different intents by activating suitable experts. It automatically and selectively activates fast DRL/heuristic-based experts that compute near-optimal planning decisions and optimal solver-based experts that are pre-configured for specific intents.

\noindent \textbf{Expert specialization.}
For each intent class, we deploy specialized optimization experts, including MIP solvers, heuristics, and learning-based experts. The details are as follows.

\textit{(1) MIP experts:} MIP solver-based experts leverage MIP solvers for $\mathbf{s}$-weighted planning optimization objectives, formulated as $\min \sum_{i=1}^M \omega_i \cdot C_i$, where $M$ is the number of objectives and $C_i$ denotes the performance value of objective $i$.
More precisely, these experts integrate solver interfaces, such as CPLEX \cite{cplex} and Gurobi \cite{gurobi}, along with decision variable generators and constraint compilers. They can provide optimal solutions for specific objectives. Note that we pre-configure some MIP solvers with typical planning optimization intents \cite{chen2020speed}. 
For example, we pre-configure an expert to jointly minimize resource overhead and latency, with equal weights assigned to both objectives, i.e., $\min \; 0.5 \cdot C_{\text{res}} + 0.5 \cdot C_{\text{lat}}$, where $C_{\text{res}}$ and $C_{\text{lat}}$ denote the resource consumption and end-to-end latency.
    
\textit{(2) Heuristic experts:} We build heuristic-based experts for fast and near-optimal network planning. 
Here, we develop experts who minimize resource overheads for node-level intent. We also build experts to minimize latency or minimize inter-device coordination overheads for link-level intents.

For node-level intents, we prioritize tasks with higher dependency levels and larger resource demands \cite{liu2024speed}. Let $\mathcal{V}$ be the set of tasks, $\ell(v)$ the dependency level of task $v$, and $r(v)$ its resource overheads. The optimization priority is defined as $\text{Priority}(v) = - a \cdot \ell(v) - b \cdot r(v)$, where $a, b > 0$ are tunable weights. Tasks are greedily placed on devices in descending order of $\text{Priority}(v)$, subject to devices' resource constraints.

For link-level intents, we aim to minimize the metadata exchanged between devices \cite{chen2024hermes}. We partition the task set $\mathcal{V}$ into disjoint subsets $\{\mathcal{V}_1, \dots, \mathcal{V}_k\}$, each of which can be placed on an individual hardware device. Let $A(u, v)$ denote the volume of inter-device metadata transferred from task $u$ to task $v$. We formulate the optimization objective as $\min \sum_{(u, v) \in \mathcal{E}_{\text{cross}}} A(u, v) \label{eq:inter_cut}$, where $\mathcal{E}_{\text{cross}} \subseteq \mathcal{E}$ is the set of edges across task subsets. 
    
\textit{(3) Learning-based experts:} We build learning-based experts based on DRL models. 
We build DRL model-based experts for network-wide intent, including maximizing task coverage by selecting task coordination paths and minimizing the number of hops within tasks by selecting task-placed devices. By learning from historical optimal decisions from MIP solvers, these experts provide near-optimal decisions while maintaining low execution time.

For example, we formulate the planning as a Markov Decision Process (MDP), where the agent observes the current planning state $s_t$ and selects an action $a_t$ (e.g., placing a task or selecting a path). This DRL model-based expert receives a reward $R_t = \lambda_1 \cdot \text{Coverage}_t - \lambda_2 \cdot \text{Hop}_t - \lambda_3 \cdot \text{Resource}_t \label{eq:reward_drl}$.
It learns a policy $\pi(a|s)$ that maximizes expected return, i.e., $\max_\pi \; \mathbf{E} \left[ \sum_{t=0}^{T} \gamma^t \cdot R_t \right] \label{eq:drl_obj}$.

\noindent \textbf{Expert profiling.} 
Each expert $E_k$ maintains a static capability vector $\mathbf{e}_k \in \mathbb{R}^{M}$, where $e_{k,i} \in [0,1]$ indicates the expert's capability on objective $i$. 
For MIP experts, $\mathbf{e}_k$ equals their pre-configured objective weights. 
For heuristic and DRL experts, $\mathbf{e}_k$ is set based on their design purpose (e.g., a latency-minimizing expert has $e_{\text{lat}}=1$ with other entries set to 0).

\noindent \textbf{Expert activation.}
Given an intent score vector $\mathbf{s}$ and expert profiles, we compute the similarity between user intent and each expert as
$\text{Sim}(E_k) = \frac{\mathbf{e}_k \cdot \mathbf{s}}{\|\mathbf{e}_k\| \cdot \|\mathbf{s}\|}$. Then, we apply a dual-channel expert activation strategy.
We summarize this activation process in Algorithm~\ref{alg:expert-selection}. 
More precisely, \sysname first takes a user intent $\mathbf{s}$ as inputs. If $\mathbf{s}$ exhibits a dominant objective (e.g., only one $\omega_i \ge \tau, \omega_i \in \mathbf{s}$), \sysname activates a single fast-response expert $E_{fast}$, i.e., heuristic or DRL-based expert, to compute near-optimal planning decisions in time (lines 1-4). Otherwise, to balance execution time and decision optimality, \sysname activates a fast expert $E_{fast}$ based on the similarity $\text{Sim}(E_k)$ between the intent score vector and each expert, while activating an MIP solver-based optimal expert $E_{MIP}$, which is configured by weights in $\mathbf{s}$ (lines 5-11). In this case, \sysname can leverage the output decisions of a fast expert to initialize decision variables in the Gurobi-based MIP solver, so that \sysname can warm-start the MIP solver for lower execution time. 

\begin{algorithm}[tb]
\small
\caption{Dual-channel Expert Activation}
\label{alg:expert-selection}
\begin{algorithmic}[1]
\Require Intent score vector $\mathbf{s}$, the set of heuristic or DRL model-based fast expert $\mathcal{E}_\text{fast}$, the MIP solver-based expert $E_\text{MIP}$, threshold $\tau$
\Ensure Activated expert set $\mathcal{E}_\text{act}$
\State $Flag_\text{high} \gets \{\omega_i \mid \omega_i \ge \tau\}$
\If{$|Flag_\text{high}| = 1$}
    \State $i^* \gets$ the element with max value in $\mathbf{s}$
    \State $\mathcal{E}_\text{act} \gets \{E_k \in \mathcal{E}_\text{fast} \mid e_{k,i^*} = 1\}$
\Else
    \For{each $E_k \in \mathcal{E}_\text{fast}$}
        \State $\text{Sim}(E_k) \gets \frac{\mathbf{e}_k \cdot \mathbf{s}}{\|\mathbf{e}_k\| \cdot \|\mathbf{s}\|}$
    \EndFor
    \State $E^*_\text{fast} \gets \arg\max\limits_{E_k \in \mathcal{E}_\text{fast}} \text{Sim}(E_k)$
    \State Configure $E_\text{MIP}$ with weights $\mathbf{s}$
    \State $\mathcal{E}_\text{act} \gets \{E^*_\text{fast}, E_\text{MIP}\}$
\EndIf
\State \Return $\mathcal{E}_\text{act}$
\end{algorithmic}
\end{algorithm}

%% file: paper/design3.tex
\section{DRL-based Expert Configuration}
\noindent \textbf{Challenges.} Existing configurations for network planning focus on single-objective or specific objective sets, suffering from low planning performance under different intents \cite{liu2024speed, chen2024eagle}.
A strawman approach is using values in the vector $\mathbf{s}$ directly as weights in solvers. However, since solvers cannot normalize performance metrics, directly applying distorts optimization outcomes as the magnitude differences of performance metrics dominate user intent.

\noindent \textbf{Key ideas.} 
In response, we pre-configure some MIP solver-based experts based on existing studies \cite{chen2020speed,chen2021mtp,chen2024hermes,zheng_dinc_2023,chen2021lightnf,chen2024eagle}. To further improve the offloading performance of pre-configured MIP solver-based experts under different user intents, we leverage DRL to configure the weights of optimization objectives within the activated expert. Here, we use the historical optimal offloading decisions from MIP solvers to train the DRL agent.

\noindent \textbf{DRL-based expert configuration.} We design a DRL agent to configure the optimization objective weights of the activated expert, where we integrate the above intent-aware scoring approach for achieving better planning performance for the given user intent.

\textbf{State space.} The state $\mathcal{S}_t$ integrates the current weight configuration with the intent score, i.e.,
\begin{equation}
\mathcal{S}_t = \langle \mathbf{s}_t \rangle
\end{equation}
where $\mathbf{s}_t = [\omega_{t,1}, \ldots, \omega_{t,M}]$ is the per-objective score vector.

\textbf{Action space.} 
Given $M$ objectives, where $M=7$, the action $a_t$ represents a learning rate adjustment factor $\text{lr}_t \in [-4, 4]$ that modulates the weight distribution. 
Here, the new objective weights $\omega_{t+1}$ are computed iteratively from the previous weights $\omega_t$ using a shift-softmax transformation, i.e., $\omega_{t+1} = \text{shift\_softmax}(\omega_t, \text{lr}_t)$, where shift-softmax is defined as:
\begin{equation}
\text{shift\_softmax}(x, h) = \begin{cases}
x & \text{if } h = 0 \\
\frac{(\exp(h \cdot x) - 1) \cdot \sum x}{\sum (\exp(h \cdot x) - 1)} & \text{otherwise}
\end{cases}
\end{equation}

\textbf{Reward function.} 
Given a weight configuration $\omega_t$, we compute the normalized improvement of each optimization objective $O_i$ at time step $t$, i.e., 
\begin{equation}
\Delta_{t, i} = d_i \cdot \frac{\text{Perf}_{t,i} - \text{Perf}_{t-1,i}}{|\text{Perf}_{t-1,i}|}
\end{equation}
where $\text{Perf}_{t,i}$ is the performance value of objective $O_i$ at time step $t$, and $d_i \in \{+1, -1\}$ is a directional factor that unifies the notion of ``improvement'' (e.g., $d_i=-1$ for latency, $d_i=+1$ for throughput). 

Then, we weight each objective's improvement by its corresponding intent score, i.e.,
\begin{equation}
\mathcal{W}_{t,i} = \omega_{t,i} \cdot \Delta_{t,i}
\end{equation}

Next, we assess the overall performance trend using the Hodges-Lehmann estimator \cite{hodges2011estimates}, which robustly captures central tendency through pairwise averages. Let $\{\mathcal{W}_{t,1}, \ldots, \mathcal{W}_{t,M}\}$ be the intent-weighted improvement samples, we compute $\mu = \text{hl\_sign}(\mathcal{W}_t) = \text{median}\left\{ \frac{\mathcal{W}_{t,i} + \mathcal{W}_{t,j}}{2} \mid 1 \leq i \leq j \leq M \right\}$. Here, $\mu$ indicates the overall trend, i.e., whether performance improves or degrades.

Finally, we define the reward $r_t$, i.e.,
\begin{equation}
r_t = \bar{\Delta}_\mathcal{W}(t) \cdot \mathbb{I}[\mu > 0] - \beta \cdot \sum_{c_j \in C} \mathbb{I}[\text{violate}(c_j)]
\end{equation}
where $\bar{\Delta}_{\mathcal{W}_t} = \sum_{i=1}^{M} \mathcal{W}_{t,i} = \sum_{i=1}^{M} \omega_{t,i} \cdot \Delta_{t,i}$ is the intent-weighted improvement, $\mathbb{I}[\text{HL\_sign}(t) > 0]$ is an indicator function that equals 1 when the trend is positive, and $\sum_{c_j \in C} \mathbb{I}[\text{violate}(c_j)]$ is a penalty for any constraint violation.

%% file: paper/evaluation.tex
\section{Case Study: ML Inference Offloading}\label{evaluation}
\noindent We evaluate \sysname through a representative case study of ML inference offloading, where ML inference tasks are offloaded onto the network of high-performance computing (HPC) hardware, e.g., Intel Tofino-series ASICs, and NVIDIA data processing units (DPUs) to achieve real-time traffic analysis. This scenario exemplifies network planning optimization with multiple competing objectives (e.g., resource utilization, latency, and throughput) under complex constraints (e.g., device capacity and link bandwidth). Further, \sysname is a domain-agnostic framework that can be extended to other network planning domains such as SFC placement and traffic engineering (see $\S$\ref{discussion}).

\subsection{Experimental Settings}

\noindent \textbf{Testbed.} We built a testbed consisting of two 12.8\,Tbps HPC hardware devices \cite{tofino} and two servers. 
The testbed is organized in a sequential topology, with the two HPC hardware devices located in the middle. The left server and the right server run MoonGen-based \cite{emmerich2015moongen} traffic senders and traffic receivers, respectively. These devices are connected via 100-Gbps links. In addition, we use another server as the control plane, running the \sysname prototype and all comparison solutions. This server is directly connected to the devices in our testbed.
Further, we leverage Mininet \cite{mininet} to model larger network topologies containing a cluster of HPC hardware. 

\noindent \textbf{Input tasks.} We build 5 types of ML inference tasks, i.e., Naive Bayes (NB), Decision Tree (DT), Random Forest (RF), SVM, and XGBoost (XGB). These tasks represent advanced ML inference workloads that require significant HPC hardware resources and need optimization when co-offloaded \cite{zheng_dinc_2023}.

\noindent \textbf{Topology.} We select two types of real-world topologies:
(1) WAN topologies from the Internet Topology Zoo \cite{knight2011internet}. 
We configure link bandwidth to 10\,Gbps.
(2) DC topologies, i.e., the Fat-Tree topology. We randomly set the link latency between 1\,us and 10\,us \cite{chen2024eagle}. 

\begin{table}[t]
\centering
\small
\caption{(Exp\#1) Latency and SRAM consumption under WAN (Internet2) topology. 
The best results are boldfaced, and the second-best results are underlined.}
\begin{tabular}{clcc}
\hline
\textbf{Intent} & \textbf{Method} & \textbf{Latency (ms)} & \textbf{SRAM} \\
\hline
  & SPD-O \cite{chen2020speed}  & \underline{327.0} & \underline{42.701\%} \\
User      & SPD-H \cite{liu2024speed}  & 4420.8 &43.376\% \\
Intent      & HMS \cite{chen2024hermes}    & 13495.0 &46.401\% \\
\#1      & DP \cite{zhang2023dapper}     & 949.0 & 43.100\% \\
      & \textbf{\sysname}    & \textbf{294.4} &\textbf{42.547\%} \\
\hline
  & SPD-O \cite{chen2020speed}  & \textbf{327.0} & \underline{42.701\%} \\
User      & SPD-H \cite{liu2024speed}  & 4420.8 &43.376\% \\
Intent      & HMS \cite{chen2024hermes}    & 13495.0 &46.401\% \\
\#2      & DP \cite{zhang2023dapper}    & 949.0 & 43.100\% \\
      & \textbf{\sysname}    & \underline{420.6} &\textbf{42.512\%} \\
\hline
  & SPD-O \cite{chen2020speed}  & \textbf{327.0}  & \underline{42.701\%} \\
User      & SPD-H \cite{liu2024speed}  & 4420.8 &43.376\% \\
Intent      & HMS \cite{chen2024hermes}    & 13495.0 &46.401\% \\
\#3      & DP \cite{zhang2023dapper}    & 949.0 & 43.100\% \\
      & \textbf{\sysname}    & \underline{382.0} &\textbf{42.490\%} \\
\hline
\end{tabular}
\label{tab:differentIntent}
\end{table}

\noindent \textbf{User intents.} We evaluate \sysname with three user intents:
(1) \textbf{U\#1}: ``Offload some ML inference tasks with high service performance''; 
(2) \textbf{U\#2}: ``Offload some tasks with higher throughput'';
(3) \textbf{U\#3}: ``In a data center, offload multiple ML tasks with low latency.''

\noindent \textbf{Baselines.} We compare \sysname with three types of solutions:
(1) MIP solver-based frameworks. We choose SPEED-OPT (SPD-O) \cite{chen2020speed}, which formulates ML inference task offloading as a mixed integer program that minimizes the weighted sum of hardware device resource overhead and end-to-end latency via Gurobi \cite{gurobi}. It computes optimal solutions under a fixed objective.
(2) Domain-specific heuristics. We combine the first-fit-by-level-and-size (FFLS) heuristic \cite{jose2015compiling} and the NodeRank heuristic \cite{liu2024speed} for minimizing device resource consumption and end-to-end latency, and name it SPEED-Heu (SPD-H).
We also choose Hermes (HMS) \cite{chen2024hermes}, which minimizes inter-device coordination overhead via a greedy heuristic.
(3) DRL-based solutions. We choose Dapper \cite{zhang2023dapper} (DP), a DRL-based SFC deployment framework. We adapt it for ML inference task offloading to minimize latency. It learns a placement policy under a fixed objective and requires retraining for new intents.

\noindent \textbf{LLM methodology.} We use GPT-4o (accessed through OpenAI's APIs) to interpret user intents. User inputs are processed as natural language text prompts. We employ carefully engineered prompts within the Dify \cite{dify2023} platform to guide structured output generation. 

\noindent \textbf{Experts.} We build an MIP solver-based expert using Gurobi \cite{gurobi}, with pre-configured weights for multiple optimization objectives. We also built two greedy-based heuristic experts that minimize inter-device coordination overhead and intra-device resource consumption, respectively. Additionally, we built two DRL model-based experts to minimize the number of hops and end-to-end latency.

We evaluate \sysname from the following research questions:
\begin{itemize}[leftmargin = *]
\item \textbf{RQ1:} How does \sysname perform on resource consumption and end-to-end performance under different tasks, topologies, and user intents?
\item \textbf{RQ2:} How does \sysname perform on execution time under different topology scales?
\item \textbf{RQ3:} How sensitive is the expert configuration of \sysname to the number of epochs?
\item \textbf{RQ4:} How much do the key modules of \sysname, i.e., dynamic expert activation and weight configuration, contribute to the overall performance?
\end{itemize}

\subsection{Comparison Experiments (RQ1)}
\noindent \textbf{(Exp\#1) \sysname achieves high performance when input intents change.} 
We evaluate the latency and device resource consumption. 
Table~\ref{tab:differentIntent} shows the offloading performance of \sysname and comparison solutions under three user intents and a WAN topology. Across user intents, \sysname achieves high performance on latency and resource consumption since it adapts to heterogeneous intents. 
Compared to baselines, \sysname reduces the latency by up to 97.8\% under a WAN (Internet2) topology. Furthermore, \sysname also achieves optimal SRAM resource consumption.

\begin{table}[t]
\centering
\small
\caption{(Exp\#2) SRAM resource (i.e., intra-device resource) overhead  under different task numbers. The best results are boldfaced, and the second-best results are underlined.}
\begin{tabular}{lccccc}
\hline
\textbf{\# of Tasks} & \textbf{20} & \textbf{40} & \textbf{60} & \textbf{80} & \textbf{100}\\
\hline
SPD-O \cite{chen2020speed}       & \underline{14.1\%} & \underline{21.4\%} &\underline{42.7\%} &\textbf{60.9\%} & \underline{71.9\%} \\
SPD-H \cite{liu2024speed}  & 14.3\% &21.9\% &43.4\% &61.9\% & \textbf{71.6\%} \\
HMS \cite{chen2024hermes}      &15.3\% &23.8\% &46.4\% &66.0\% & 76.3\%\\
DP \cite{zhang2023dapper}       & \underline{14.1\%} &23.0\% & 43.1\% &62.3\% & 79.2\%\\
\textbf{\sysname}    &\textbf{13.9\%} &\textbf{21.3\%} &\textbf{42.6\%} &\underline{61.0\%} & 74.0\%\\
\hline
\end{tabular}
\normalsize
\label{tab:sram}
\end{table}

\begin{table}[t]
\centering
\small
\caption{(Exp\#3) PHV resource (i.e., inter-device coordination) overhead under different task numbers. The best results are boldfaced, and the second-best results are underlined.}
\begin{tabular}{lccccccc}
\hline
\textbf{\# of Tasks} & \textbf{20} & \textbf{40} & \textbf{60} & \textbf{80} & \textbf{100}\\
\hline
SPD-O \cite{chen2020speed}      &\underline{13.2\%} &\textbf{25.0\%} &\underline{40.9\%} &\underline{53.2\%} & \textbf{68.9\%} \\
SPD-H \cite{liu2024speed} &13.6\% &26.0\% &42.0\% &54.8\% &\underline{69.3\%} \\
HMS \cite{chen2024hermes}      &14.9\% &28.6\% &46.1\% &60.0\% &75.6\%\\
DP \cite{zhang2023dapper}   & 13.8\% &\underline{25.2\%} &41.6\% &53.9\% & 72.7\%\\
\textbf{\sysname}    &\textbf{13.1\%} &\textbf{25.0\%} &\textbf{40.7\%} &\textbf{53.1\%} &72.4\%\\
\hline
\end{tabular}

\label{tab:phv}
\end{table}

\begin{table}[t]
\centering
\small
\caption{(Exp\#4) Latency across topologies and input task numbers. The ``T\#1'' represents the Abovenet topology, ``T\#2'' is the Internet2 topology, and ``T\#3'' is a small FatTree topology. ``T\#1'' and ``T\#2'' are WAN topologies. ``T\#3'' is a DC topology. 
``N/A'' means all tasks are deployed on a \textit{single device}, where the inter-device latency can be negligible. 
The best results are boldfaced, and the second-best are underlined.}
\small
\begin{tabular}{llccccc}
\hline
\textbf{Topo} & \textbf{\# of Tasks} & \textbf{20} & \textbf{40} & \textbf{60} & \textbf{80} & \textbf{100} \\
\hline
T\#1 & SPD-O \cite{chen2020speed}      & \textbf{N/A} &\underline{11.6} &\textbf{N/A} &\textbf{N/A} &\underline{62.0}\\
Lat.   & SPD-H \cite{liu2024speed}      & \underline{1026.8} &2119.4 &\underline{2947.4}& 3970.6 &5033.6\\
(ms)   & HMS \cite{chen2024hermes}        & 1147.0 &3480.0 &8250.0 &10884 &16020\\
   & DP \cite{zhang2023dapper}         & \textbf{N/A} &\textbf{N/A} &\textbf{N/A} &\underline{619.0} &2089.0\\
   & \sysname         & \textbf{N/A} &\textbf{N/A} &\textbf{N/A} &\textbf{N/A} &\textbf{N/A}\\
\hline
T\#2 & SPD-O \cite{chen2020speed}      & \textbf{66.0} &\textbf{79.0} &\textbf{327.0} &\textbf{782.0} &\textbf{2889.0}\\
Lat.   & SPD-H \cite{liu2024speed}      & 1133.6 &3045.2 &4420.8 &6977.4 &8741.2\\
(ms)   & HMS \cite{chen2024hermes}        & 2178.0 &6785.0 &13495 &20816 &29200\\
   & DP \cite{zhang2023dapper}         &143.0 &637.0& 949.0& 2243.0 &8361.0\\
   & \sysname        &\underline{108.0} &\underline{123.0} &\underline{459.0} &\underline{1371.0} &\underline{4129.0}\\
\hline
T\#3 & SPD-O \cite{chen2020speed}      & \textbf{N/A} &\textbf{N/A} &0.4 &\textbf{N/A} &\underline{0.3}\\
Lat.   & SPD-H \cite{liu2024speed}      & 120.3 &571.6 &812.2& 1252.5& 1922.5\\
(us)   & HMS \cite{chen2024hermes}         & \underline{42.3} &\underline{176.7} &\underline{463.5} &864.3 &1280.6\\
   & DP \cite{zhang2023dapper}         &\textbf{N/A}&\textbf{N/A}&\textbf{N/A}&\underline{36.3} &137.5\\
   &  \sysname       & \textbf{N/A} &\textbf{N/A} &\textbf{N/A} &\textbf{N/A} &\textbf{N/A}\\
\hline
\end{tabular}
\label{tab:latency}
\end{table}

\noindent \textbf{(Exp\#2) \sysname achieves low intra-device resource overheads.} We quantify the intra-device overhead via SRAM resource consumption. More precisely, we vary the input topologies and task numbers. We fix user intent as ``offloading more ML inference tasks.'' Then, we measure the SRAM resource consumption of \sysname and baselines. 
In Table~\ref{tab:sram}, \sysname consumes fewer SRAM resources, reducing SRAM resources by up to 6.6\% since we build and activate experts for minimizing intra-device overhead.

\noindent \textbf{(Exp\#3) \sysname achieves low inter-device coordination overheads.} We quantify the inter-device overhead by measuring packet header vector (PHV) resource consumption. More precisely, we fix the user intent the same as Exp\#2. Then, we measure the PHV resource consumption of \sysname and baselines.
In Table~\ref{tab:phv}, \sysname achieves near-optimal offloading, reducing PHV resource overhead by up to 11.5\%.

\noindent \textbf{(Exp\#4) \sysname achieves high end-to-end performance.} We evaluate the end-to-end performance of \sysname and baselines. We measure the latency under different input task numbers and topologies. We fix the user intent as ``offloading more ML inference tasks while preserving low latency.''
Table~\ref{tab:latency} shows the latency of \sysname and baselines.
\sysname achieves the near-optimal latency across topologies and task numbers. 

\noindent \textbf{(Exp\#5) DL inference workloads.}
We evaluate \sysname on two additional DL inference workloads, i.e., YOLO-v5s and BERT-base, and measure SRAM resource consumption and throughput. 
Table~\ref{tab:dl-workloads} shows that \sysname achieves high throughput while maintaining competitive resource consumption on both workloads.
 
\begin{table}[t]
\small
\caption{(Exp\#5) Performance on DL inference workloads. The best results are \textbf{boldfaced}, and the second-best results are underlined.}
\label{tab:dl-workloads}
\begin{tabular}{llcc}
\toprule
\textbf{Workload} & \textbf{Method} & \textbf{SRAM (\%)} & \textbf{Thpt.\ (Gbps)} \\
\midrule
\multirow{5}{*}{YOLO-v5s}
 & SPD-O~\cite{chen2020speed} & 16.214 & 2 \\
 & SPD-H~\cite{liu2024speed} & 16.227 & 7 \\
 & HMS~\cite{chen2024hermes} & 16.178 & 2 \\
 & DP~\cite{zhang2023dapper} & \underline{16.127} & \textbf{8} \\
 & \textbf{OmniPlan} & \textbf{16.067} & \textbf{8} \\
\midrule
\multirow{5}{*}{BERT-base}
 & SPD-O~\cite{chen2020speed} & \textbf{21.365} & 1 \\
 & SPD-H~\cite{liu2024speed} & 21.425 & 6 \\
 & HMS~\cite{chen2024hermes} & 21.420 & 2 \\
 & DP~\cite{zhang2023dapper} & 21.560 & \textbf{8} \\
 & \textbf{OmniPlan} & \underline{21.455} & \textbf{8} \\
\bottomrule
\end{tabular}
\end{table}

\subsection{Execution Time (RQ2)}
\noindent \textbf{(Exp\#6) \sysname achieves low execution time.} We measure the execution time of \sysname against baselines under topologies with different device numbers. 
\sysname achieves fast offloading by activating DRL or heuristic-based experts and achieves optimal offloading by activating solver-based experts. Thus, we measure the two execution time of \sysname. 
Table~\ref{tab:execution_time} indicates that in the fast offloading stage, \sysname takes less than 0.1 seconds to compute the offloading decisions, which is competitive to existing fast offloading solutions \cite{chen2024hermes,liu2024speed,zhang2023dapper}. In the optimal offloading stage, \sysname also takes a shorter execution time than SPD-O \cite{chen2020speed} in most of cases, reducing execution time by up to 59.2\%.

\begin{table}[t]
\centering
\small
\caption{(Exp\#6) Execution time (s) with a varying number of hardware devices (HD). ``Fast'' means achieving fast offloading and ``Opt.'' means achieving optimal offloading. \sysname achieves fast offloading by activating DRL or heuristic-based experts and achieves optimal offloading by activating solver-based experts. The best results are boldfaced.}
\begin{tabular}{clcccc}
\hline
 &\textbf{Method} & \textbf{8 HD} & \textbf{16 HD} & \textbf{24 HD} & \textbf{32 HD}\\
\hline
Fast &SPD-H \cite{liu2024speed}   &0.026 &0.028 &0.029 &0.039 \\
 & HMS \cite{chen2024hermes}     &\textbf{0.013}& \textbf{0.022}& \textbf{0.028} &\textbf{0.038}\\
&DP \cite{zhang2023dapper}   &4.519 &4.957 &4.633 &4.435 \\
&\sysname    &\textbf{0.013}& \textbf{0.022}& \textbf{0.028}& \textbf{0.038} \\
\hline
Opt. &SPD-O \cite{chen2020speed}       &170.4 &601.6 &602.5 &602.9 \\
  &\sysname &\textbf{78.3}& \textbf{245.7}& \textbf{379.0} &\textbf{435.0} \\
\hline
\end{tabular}
\label{tab:execution_time}
\end{table}

\begin{table}[t]
\centering
\small
\caption{(Exp\#7) Performance of \sysname in different configuration epochs. In the 6th epoch, the expert weight configuration has achieved the optimum, and \sysname automatically stops at the 7th epoch.}
\label{tab:epochNumer}
\setlength{\tabcolsep}{4pt}
\resizebox{\columnwidth}{!}{%
\begin{tabular}{lcccccc|c}
\toprule
 & \multicolumn{6}{c|}{\textbf{Expert weights configuring}} & \textbf{Optimum} \\
\cmidrule(lr){2-7} \cmidrule(lr){8-8}
\textbf{Epoch ID} & \textbf{0} & \textbf{1} & \textbf{2} & \textbf{3} & \textbf{4} & \textbf{5} & \textbf{6 \& 7} \\
\midrule
SRAM & 42.6\% & 42.4\% & 42.6\% & 42.5\% & 42.4\% & 42.8\% & \textbf{42.4\%} \\
Latency (ms) & 691 & 310 & 310 & 509 & 255 & 314 & \textbf{258} \\
Thpt. (Gbps) & 3.0 & 9.0 & 9.0 & 6.0 & 7.0 & 9.0 & \textbf{9.0} \\
\bottomrule
\end{tabular}%
}
\end{table}

\begin{table}[t]
\small
\setlength{\tabcolsep}{4pt} 
\caption{(Exp\#8) Impact of dynamic expert configuration.} 
\label{tab:human-expert}
\begin{tabular}{llccc}
\hline
\textbf{Intent type} & \textbf{Method} & \textbf{Lat.\ (ms)} & \textbf{Thpt.\ (Gbps)} & \textbf{SRAM (\%)} \\
\hline
\multirow{2}{*}{1} & Human-expert & 358 & 6 & 42.636 \\
 & OmniPlan & 310 & 9 & 42.395 \\
\hline
\multirow{2}{*}{2} & Human-expert & 329 & 7 & 42.656 \\
 & OmniPlan & 310 & 9 & 42.401 \\
\hline
\multirow{2}{*}{3} & Human-expert & 404 & 5 & 42.608 \\
 & OmniPlan & 223 & 7 & 42.564 \\
\hline
\end{tabular}
\end{table}

\begin{table}[t]
\centering
\small
\caption{(Exp\#9) Impact of dynamic optimization objective weights on latency, hardware SRAM resource consumption, and throughput (i.e., Thpt.). ``Equal'' indicates that three optimization objectives are equally weighted. ``Latency'' means the weight for minimizing latency is set to 50\%, while the other two objectives are weighted at 25\% each. ``Throughput'' and ``Resource'' use the same pattern. The best results are boldfaced, and the second-best results are underlined.}
\begin{tabular}{lccc}
\hline
\textbf{Method} & \textbf{Latency (ms)} & \textbf{SRAM} & \textbf{Thpt. (Gbps)} \\
\hline
Equal      &883.5 &42.568\%	&2.1\\
Latency    & \underline{319.2} & 42.478\% &3.0\\
Throughput & 361.8 & \textbf{42.437\%} &\underline{4.0}\\
Resource      & 502.8 &\underline{42.591\%} &\textbf{6.0}\\
\textbf{\sysname}   & \textbf{310.0} &42.534\% &\textbf{6.0}\\
\hline
\end{tabular}
\label{tab:weighting_strategies}
\end{table}

\subsection{Sensitivity to Epoch Number (RQ3)}

\noindent \textbf{(Exp\#7) Impact of epoch number on DRL-based expert configuration.} 
We take the user intent, i.e., ``I want to maximize the bandwidth with the highest priority, and then minimize the latency and SRAM usage as possible'', Internet2 topology, and 60 tasks as input. In the 6th epoch, \sysname achieves optimal offloading. We measure the latency, throughput, and SRAM resource consumption of \sysname in each epoch. Table~\ref{tab:epochNumer} illustrates that compared to the original expert weights, the expert configuration of \sysname improves throughput by 3$\times$ and reduces the latency, while maintaining low SRAM resource consumption. Moreover, \sysname achieves the optimal configuration in epoch 6, which means that \sysname only requires a few configuration epochs. We put the weight values of each epoch in our appendix.

\subsection{Ablation Analysis (RQ4)}
\noindent \textbf{(Exp\#8) Impact of dynamic expert configuration.}
We compare \sysname with human-expert-tuned baselines. We invite 3 graduate researchers with network optimization experience to manually tune MIP objective weights for three additional types of intents beyond U\#1-U\#3. 
Each is given the MIP formulation, Gurobi, Internet2 topology with 60 tasks, and intent descriptions, with up to 10 weight configurations per intent. We report the best result among the 3 researchers.
In Table~\ref{tab:human-expert}, \sysname outperforms baselines across intents and metrics thanks to the DRL-based configuration.

\noindent \textbf{(Exp\#9) Impact of dynamic weights.} We evaluate the impact of dynamic expert weights in \sysname.
More precisely, we take these optimization objectives, including minimizing latency, maximizing throughput, and minimizing HPC hardware device resources. We build four static weighted-based strawman solutions using Gurobi \cite{gurobi}, including (1) equal weights for all objectives, i.e., ``Equal'', (2) assigning 50\% of the weight to an objective, with the remaining two objectives sharing the rest equally, i.e., ``Latency'', ``Throughput'', and ``Resource''. We take the user intents, i.e., U\#1 to U\#3, as our inputs for each objective.
In Table~\ref{tab:weighting_strategies}, compared with static weighted-based groups, \sysname achieves the lower latency, which matches user intent U\#1, i.e., ``offload some ML inference tasks, where I need a high service performance'' better. Also, for another two metrics, our results show a similar trend.

\noindent \textbf{(Exp\#10) LLM-based translation.}
We compare the LLM-based intent translation in \sysname with three baselines on 50 diverse intents covering all 7 objectives, i.e., rule-based keyword matching, template-based matching with 20 regex templates, and a fine-tuned BERT classifier trained on 100 labeled pairs.
We evaluate intent translation accuracy and expert activation accuracy. 
In Table~\ref{tab:llm-alternatives}, \sysname achieves the highest translation and activation accuracy.
 
\begin{table}[t]
\small
\caption{(Exp\#10) Comparison of intent translation methods on 50 diverse intents. The best results are \textbf{boldfaced}.}
\label{tab:llm-alternatives}
\begin{tabular}{lcc}
\toprule
\textbf{Method} & \textbf{Translation Acc.} & \textbf{Activation Acc.} \\
\midrule
Rule-based & 72\% & 86\% \\
Template-based & 22\% & 34\% \\
Fine-tuned classifier & 78\% & 70\% \\
LLM-based (Ours) & \textbf{82\%} & \textbf{92\%} \\
\bottomrule
\end{tabular}
\end{table}

\noindent \textbf{(Exp\#11) Robustness of expert activation.}
We evaluate the robustness of expert activation in \sysname across 5 WAN topologies and 2 DC topologies with 3 types of intent. \sysname activates the same fast expert for each intent across all 7 topologies. 
We further generate 10 paraphrases per intent, e.g., ``minimize delay'', ``fast response'', and ``latency is critical''. \sysname activates the same expert for each intent across paraphrases. 
Table~\ref{tab:robustness} reports the cosine similarity between paraphrased and original intent vectors. The gap between the selected and second-best expert $>0.15$ in all cases.

\begin{table}[t]
\small
\caption{(Exp\#11) Robustness of expert activation}
\label{tab:robustness}
\begin{tabular}{lccc}
\toprule
\textbf{Intent type} & \textbf{\# Paraphrases} & \textbf{Min Cos-Sim} & \textbf{Avg Cos-Sim} \\
\midrule
1 & 10 & 0.811 & 0.965 \\
2 & 10 & 0.711 & 0.910 \\
3 & 10 & 0.707 & 0.877 \\
\bottomrule
\end{tabular}
\end{table}

%% file: paper/discussion.tex
\section{Discussion}\label{discussion}
\noindent \textbf{Fallback for out-of-scope intents.}
For intents that involve objectives outside the current 7-dimensional space, \sysname applies a three-step fallback procedure. First, it flags the out-of-scope objective and maps it to the semantically closest existing dimension. Second, it notifies the user and presents two options, i.e., proceeding with the approximate mapping or registering a custom expert for the new objective. Third, it executes the chosen option, where newly registered experts are persisted for future planning requests.

%% file: paper/limitations.tex
\section{Limitations and Ethical Considerations}
\noindent \textbf{Limitations and future works.}
\sysname relies on LLMs for intent interpretation, which may introduce microsecond to seconds of latency from API calls and potential hallucination in understanding complex or ambiguous intents.
In the future, we plan to mitigate hallucinations via structured output enforcement and semantic validation against predefined anchors, with automatic re-query for non-conforming outputs. We also plan to replace the API-based LLMs with fine-tuned local LLMs to reduce latency. We plan to use multi-vector representations to support more complex intent.

\noindent \textbf{Ethical Considerations.} 
(1) Data privacy. Datasets and topologies used in this work are publicly available and user intents are synthetic and do not involve real user data.
(2) Potential misuse.
While \sysname is designed for legitimate network optimization, automated planning tools could theoretically be adapted for malicious purposes, such as optimizing attack traffic distribution.
We encourage operators deploying such systems to implement appropriate access controls and audit mechanisms.
(3) Consent and bias. This work does not involve human subjects, personally identifiable information, or decisions about individuals.
Therefore, concerns regarding consent and bias toward specific demographic groups are not applicable.

%% file: paper/conclusion.tex
\section{Conclusion}\label{conclusion}
\noindent In this paper, we have proposed \sysname, an adaptive framework for timely and near-optimal network planning optimization. Its key idea is to leverage LLMs to understand user intents and activate suitable experts within the MoE architecture to achieve intent-adaptive planning. 
We evaluate \sysname on a representative case study, i.e., distributed machine learning (ML), where we leverage it to offload a wide spectrum of ML inference tasks onto a network of hardware devices. Our experiments on a real-world testbed have clearly demonstrated that \sysname achieves near-optimal and low-execution-time offloading for real-world ML inference tasks.

%% file: paper/appendix.tex
\section{Appendix}

\subsection{Notation of Main Symbols}

Table~\ref{tab:notation} summarizes the notation of main symbols in this paper.

\begin{table}[b]
\caption{Notation of main symbols.}
\small
\setlength{\tabcolsep}{2pt}            
\renewcommand{\arraystretch}{1}     
\begin{tabular}{p{0.30\columnwidth} p{0.64\columnwidth}}
\toprule
\multicolumn{2}{c}{Symbols of Intent Translation}\\
\midrule
$u$ & The user intents in natural language. \\
$\mathcal{T}$ & The standardized textual representations. \\
$\mathbf{s}$ & Preference vector of user.\\
$M$ & Number of optimization objectives.\\
$\tau_{\mathrm{net}}$ & Domain prior in prompts.\\
$\tau_{\mathrm{format}}$ & Output-format prior in prompts.\\
$\omega_{\text{res}}, \omega_{\text{pwr}}, \omega_{\text{thr}}, \omega_{\text{lat}}$ & Quantified value of user preference. \\ 
$\omega_{\text{coor}}, \omega_{\text{hop}}, \omega_{\text{cover}}$ & \\
$K$ & Number of intent anchors.\\
$S_i$ & Intent anchor at level $i$.\\
$\mathbf{p}_i$ & Embedding of $S_i$ via sentence-BERT.\\
$\theta$ & curve parameter \\
$\mathcal{L}(\theta)$ & Cubic-spline intent curve.\\
$L_{i,i+1}$ & Arc-length between anchors $i$ and $i{+}1$.\\
$\phi_{i,i+1}$ & Arc-length weight of segment $(i,i{+}1)$.\\

\midrule
\multicolumn{2}{c}{Symbols of Expert Activation}\\
\midrule

$C_i$ & Performance value of objective $i$.\\
$\mathcal{V}$ & Set of tasks.\\
$\ell(v)$ & Dependency level of task $v$.\\
$r(v)$ & Resource overheads of task $v$.\\
$a, b$ & Tunable weights for task priority.\\
$A(u, v)$ & Volume of inter-metadata from task $u$ to $v$.\\
$\mathcal{E}_{\text{cross}}$ & Set of edges across task subsets.\\
$R_t$ & Reward received at step $t$.\\
$\pi(a|s)$ & Learned policy maximizing expected return.\\
$E_k$ & Expert $k$.\\
$\mathbf{e}_k$ & Capability vector of expert $k$.\\
$\text{Sim}(E_k)$ & Similarity between user intent and expert $k$.\\
$\tau$ & Threshold for activating a single expert.\\
$Flag_{\text{high}}$ & Set of objectives with weights exceeding $\tau$.\\
$\mathcal{E}_{\text{fast}}$ & Set of heuristic or DRL-based fast experts.\\
$E_{\text{MIP}}$ & MIP solver-based expert.\\
$\mathcal{E}_{\text{act}}$ & Activated expert set.\\

\midrule
\multicolumn{2}{c}{Symbols of Expert Configuration}\\
\midrule

$\mathcal{S}_t$ & State of the DRL agent at time step $t$.\\
$a_t$ & Action selected at time step $t$.\\
$\text{lr}_t$ & Learning rate adjustment factor at time step $t$.\\
$\Delta_{t, i}$ & Normalized improvement of objective $i$.\\
$d_i$ & Directional factor for objective $i$.\\
$\text{Perf}_{t,i}$ & Performance value of objective $i$ at time step $t$.\\
$\mathcal{W}_{t,i}$ & Intent-weighted improvement sample of objective $i$ at time step $t$.\\
$\mu$ & Overall performance trend indicator (Hodges-Lehmann estimator).\\
$r_t$ & Reward received at time step $t$.\\
$\bar{\Delta}_{\mathcal{W}_t}$ & Total intent-weighted improvement at step $t$.\\
$\beta$ & Penalty coefficient for constraint violations.\\
$C$ & Set of constraints.\\
$c_j$ & Specific constraint in set $C$.\\

\bottomrule
\end{tabular}
\label{tab:notation}
\end{table}

\subsection{Design Details}
 
\noindent \textbf{Initial weight perturbation in expert configuration.} When user intents contain conjunctions (e.g., ``maximize bandwidth and minimize latency''), the LLM often assigns identical semantic scores to multiple objectives, resulting in equal initial weights. In this case, since the shift-softmax function cannot differentiate identical values, we introduce small random perturbations to break symmetry:
\begin{equation}
\omega_0 = \omega_{\text{init}} + \epsilon, \quad \epsilon \sim \mathcal{U}(-0.05, 0.05)
\end{equation}
where $\omega_{\text{init}}$ is the initial weight vector from intent-aware scoring.

\noindent \textbf{Recursive weight refinement in expert configuration.} When weight adjustments become marginal (i.e., $\max_i |\omega_{t+1,i} - \omega_{t,i}| < \tau$ where $\tau = 0.01$), the DRL agent enters a recursive refinement phase. In this case, shift-softmax cannot further differentiate objectives with identical or near-identical weights, which is a common scenario when LLMs assign equal scores to conjoined objectives in user intents. 
In response, we execute the recursive procedure that fixes the dominant weight corresponding to $\arg\max_i \omega_t$ and recursively optimizes the remaining weight subspace with renewed perturbations. We continually execute this process until all weights are sufficiently differentiated to achieve fine-grained alignment with user preferences.

\begin{table*}[ht]
\centering
\small
\caption{(Exp\#13) Weights of optimization objectives inside the activated expert change. Note that the weights of the other optimization objectives equal 0. Thus, we omit them. By the sixth epoch, the expert weight configuration has reached the optimum, where any weight changes will decrease the intent-aware score. We show weights from epochs 0 to 10, but in actual runs, \sysname will automatically stop at the seventh epoch.}
\label{tab:weightChanges}
\begin{tabular}{lcccccc|ccccc}
\toprule
 & \multicolumn{6}{c|}{\textbf{Expert weights configuring}} & \multicolumn{5}{c}{\textbf{Achieved the optimal configuration}} \\
\cmidrule(lr){2-7} \cmidrule(lr){8-12}
\textbf{Epoch ID} & \textbf{0} & \textbf{1} & \textbf{2} & \textbf{3} & \textbf{4} & \textbf{5} & \textbf{6} & \textbf{7} & \textbf{8} & \textbf{9} & \textbf{10} \\
\midrule
Weight of minimizing SRAMs   & 0.1775 & 0.2619 & 0.3020 & 0.3191 & 0.3136 & 0.3059 & 0.2210 & 0.2210 & 0.2210 & 0.2210 & 0.2210 \\
Weight of minimizing latency & 0.1775 & 0.2619 & 0.3020 & 0.3191 & 0.3136 & 0.3059 & 0.3907 & 0.3907 & 0.3907 & 0.3907 & 0.3907 \\
Weight of maximizing throughput & 0.6450 & 0.4762 & 0.3959 & 0.3617 & 0.3727 & 0.3882 & 0.3882 & 0.3882 & 0.3882 & 0.3882 & 0.3882 \\
\bottomrule
\end{tabular}
\end{table*}

\subsection{More Experimental Results}
\begin{figure}[t]
    \centering
    \includegraphics[width=0.75\linewidth]{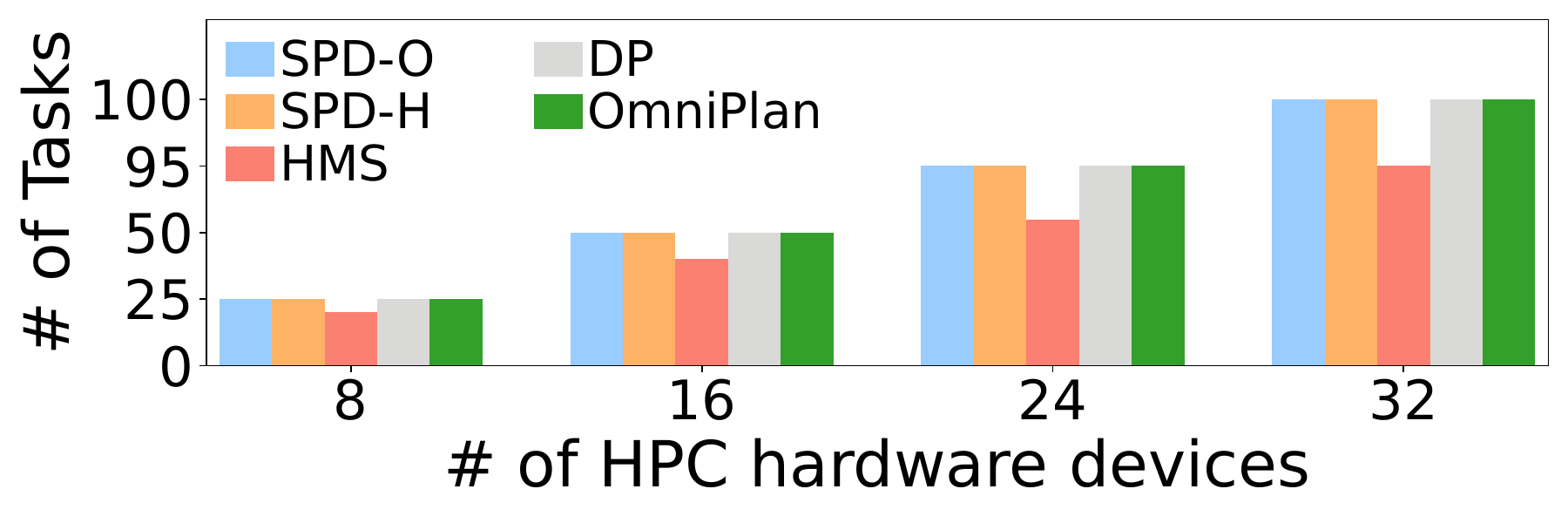}
    \caption{(Exp\#12) Number of successfully offloaded tasks with a varying number of HPC hardware devices.}
    \label{fig:taskNumber}
\end{figure}

\noindent \textbf{(Exp\#12) \sysname offloads more tasks.} We evaluate the number of successfully offloaded tasks in \sysname and comparison solutions. More precisely, we randomly select 200 ML tasks. Then, we take these ML tasks as input and offload them into topologies with different hardware device numbers, ranging from 8 to 32 devices. Compared to HMS, \sysname offloads by up to 1.3$\times$ tasks. 
Figure~\ref{fig:taskNumber} shows that \sysname can offload more ML tasks. 

\noindent \textbf{(Exp\#13) Weight changes during expert configuration.}
Table~\ref{tab:weightChanges} shows the weights of optimization objectives in the activated experts. We can see that as the epoch number increases, \sysname fine-tunes these weights and achieves the optimal weight configuration in the sixth epoch. 

\subsection{More Discussion}
\label{moreDiscussion}

\noindent \textbf{Extend to more network planning scenarios.} 
We can extend \sysname to many network planning scenarios, such as SFC placement, vehicle path scheduling, and traffic engineering, where we can define domain-specific optimization objectives, map them to the intent vector, and construct the corresponding experts. 
More precisely, since most network planning problems share common objectives, such as latency, throughput, and resource consumption, the existing intent vector structure can be reused with minor adaptations. For expert construction, existing approaches from the literature can be directly integrated as experts within \sysname. 

\noindent \textbf{Generalizability of \sysname.} \sysname is a general-purpose framework for intent-aware network planning optimization, where its architecture does not rely on domain-specific semantics. 
More precisely, the LLM-based intent translation operates purely on natural language and outputs a normalized preference vector encoding user priorities over optimization objectives. 
The MoE-based expert activation (i.e., Algorithm~1) selects experts based on cosine similarity between the preference vector and expert capability vectors, without interpreting the underlying domain meaning. 
Similarly, the DRL-based weight configuration adjusts objective weights based solely on performance feedback signals, where \sysname treats the optimization objectives as abstract dimensions. 
Therefore, these designs ensure that \sysname is domain-agnostic.

\noindent \textbf{Engineering effort.} 
\sysname reduces engineering effort in three ways. First, existing optimization algorithms can be wrapped as experts via a standard interface. 
Second, the expert library is extensible since adding a new expert only requires registering a capability vector $\mathbf{e}_k$ and its optimization logic. 
Third, DRL-based expert configuration automates weight tuning. In our prototype, the entire framework contains 4,105 lines of code, and the DRL configuration module converges within approximately 15 minutes.

\subsection{MIP Formulation for ML Offloading}
\label{app:mip}

We present the complete MIP formulation used in our case study. 

\noindent \textbf{Input and output.} \sysname takes (1) a set of ML inference tasks, where each task is decomposed into multiple match-action tables (MATs), i.e., $\mathcal{M}$, with dependency relations $\mathcal{E}$, where $e=(u,v)$ means $v$ needs the output of $u$, (2) a network topology containing device set $\mathcal{D}$ and link set $\mathcal{L}$, with device resource capacities and link bandwidths, and (3) user intent weights $\{\omega_i\}$ derived from the intent translation module as input.
\sysname outputs (1) $x_{v,d} \in \{0,1\}$, where it equals 1 if MAT $v$ is deployed on device $d$, and 0 otherwise, (2) $y_{e,l} \in \{0,1\}$, where it equals 1 if dependency edge $e$ uses link $l$ for inter-device communication, and 0 otherwise.

\noindent \textbf{Objective function.} 
We formulate an optimization as:
\begin{equation}
\min \sum_{i=1}^{M} \omega_i \cdot C_i
\label{eq:mip_obj}
\end{equation}
where $M$ is the number of objectives, $\omega_i$ is the weight for objective $i$ derived from user intent and $C_i$ is its performance value.

\noindent \textbf{Constraints.} We identify the following constraints.

\textit{(1) MAT assignment.} Each MAT must be deployed on exactly one device, i.e.,
\begin{equation}
\sum_{d \in \mathcal{D}} x_{v,d} = 1, \quad \forall v \in \mathcal{M}
\label{eq:c1}
\end{equation}

\textit{(2) Device resource capacity.} The total resource consumption of all MATs deployed on a device must not exceed its capacity, i.e.,
\begin{align}
\sum_{v \in \mathcal{M}} r_v^{(\text{SRAM})} \cdot x_{v,d} &\leq R_d^{(\text{SRAM})}, \quad \forall d \in \mathcal{D} \label{eq:c2a} \\
\sum_{v \in \mathcal{M}} r_v^{(\text{PHV})} \cdot x_{v,d} &\leq R_d^{(\text{PHV})}, \quad \forall d \in \mathcal{D} \label{eq:c2b}
\end{align}
where $r_v$ and $R_d$ denote the SRAM/PHV requirement of MAT $v$ and capacity of device $d$, respectively.

\textit{(3) Link bandwidth capacity.} The total data transfer on a link must not exceed its bandwidth capacity, i.e.,
\begin{equation}
\sum_{e=(u,v) \in \mathcal{E}} \hat{A}(u,v) \cdot y_{e,l} \leq B_l, \quad \forall l \in \mathcal{L}
\label{eq:c3}
\end{equation}
where $\hat{A}(u,v)$ denotes the data volume transferred from MAT $u$ to MAT $v$ along dependency edge $e=(u,v)$, and $B_l$ is the bandwidth of link $l$.

\textit{(4) Flow conservation.} For each dependency edge, the routing path must form a valid path from the source device to the destination device, i.e.,
\begin{equation}
\sum_{l \in \delta^+(d)} y_{e,l} - \sum_{l \in \delta^-(d)} y_{e,l} = 
\begin{cases}
1 & \text{if } d = \text{src}(e) \\
-1 & \text{if } d = \text{dst}(e) \\
0 & \text{otherwise}
\end{cases}
\quad \forall e \in \mathcal{E},\ \forall d \in \mathcal{D}
\label{eq:c4}
\end{equation}

\textit{(5) MAT dependency.} An ML inference task is decomposed into multiple MATs with data dependencies. If MAT $u$ is a predecessor of MAT $v$ (i.e., $v$ requires the output of $u$), they must be deployed on the same device or connected via a valid routing path. Furthermore, when deployed on the same device, $u$ must be placed in an earlier pipeline stage than $v$.

\noindent \textbf{Solver.} Following established practices in network function offloading optimization~\cite{liu2024speed,chen2024hermes,zheng_dinc_2023,chen2021mtp}, we employ Gurobi~\cite{gurobi} as the standard MILP solver to compute optimal deployment decisions.